\documentstyle[preprint,eqsecnum,aps,amssymb]{revtex}

\begin{document}

\newcommand{\be}{\beta}
\newcommand{\la}{\lambda}
\newcommand{\ep}{\epsilon}
\newcommand{\fr}{\frac}
\newcommand{\cam}{{\cal M}}
\newcommand{\caz}{{\cal Z}}
\newcommand{\cao}{{\cal O}}
\newcommand{\cac}{{\cal C}}
\newcommand{\cah}{{\cal H}}
\newcommand{\pnenj}{\prod_{j=1}^p\nu_j}
\newcommand{\nn}{\nonumber}
\newcommand{\intsi}{\int\limits_{\Sigma}d\sigma_x\,\,}
\newcommand{\back}{\bar{\Phi}}
\newcommand{\coba}{\bar{\Phi}^{\dagger}}
\newcommand{\abl}{\partial}
\newcommand{\qpi}{(4\pi)^{\frac{q+1} 2}}
\newcommand{\sul}{\sum_{l=-\infty}^{\infty}}
\newcommand{\snenp}{\sum_{n_1,...,n_p=0}^{\infty}}
\newcommand{\tint}{\int\limits_0^{\infty}dt\,\,}
\def\bea{\begin{eqnarray}}
\def\eea{\end{eqnarray}}
\def\be{\begin{equation}}
\def\ee{\end{equation}}


\draft
\title{Simple criterion for the occurrence of Bose-Einstein condensation and
the Meissner-Ochsenfeld effect}
\author{
Klaus Kirsten\thanks{E-mail address:
{\tt kirsten@tph100.physik.uni-leipzig.de}}}
\address{Universit{\"a}t Leipzig, Institut f{\"u}r Theoretische Physik,\\
Augustusplatz 10, 04109 Leipzig, Germany}
\author{David J.~Toms\thanks{E-mail address:
{\tt d.j.toms@newcastle.ac.uk}}}
\address{Department of Physics, University of Newcastle Upon Tyne,\\
Newcastle Upon Tyne, United Kingdom NE1 7RU}

\date{\today}
\maketitle
\begin{abstract}
We examine the occurrence of Bose-Einstein condensation in both 
nonre\-lativistic
and relativistic systems with no self-interactions in a general setting. A 
simple condition for the occurrence of Bose-Einstein condensation is given.
We show that condensation can occur only if $q\geq 3$, where $q$ is the 
dimension associated with the continuous part of the eigenvalue spectrum of 
the Hamiltonian for nonrelativistic systems or the spatial part of the 
Klein-Gordon operator for relativistic systems. Furthermore we show
that the criterion for the appearance of the Meissner-Ochsenfeld effect is
closely connected with that 
for the appearance of Bose-Einstein condensation.
\end{abstract}
\narrowtext
\section{Introduction}
One of the most interesting properties of a system of bosons is that under
certain conditions it is possible to have a phase transition at a critical
value of the temperature in which a macroscopic fraction of the bosons can condense into the
ground state. This was first predicted over 70 years ago for the ideal 
nonrelativistic Bose gas \cite{bose24,einsteinpreus24} and is nowadays
well known to happen if the spatial dimension $D\geq 3$. (See 
\cite{landaulifshitz69} for the case $D=3$ and \cite{may64} for general $D$.) The analogous phenomenon for ideal relativistic gases has also been 
studied and the criterion on the spatial dimension has been found to be the
same as that for nonrelativistic gases \cite{kapusta81,haberweldon81,haberweldon82a}.

Of course it is of great interest to see what influence different types of
interactions have on the condensation. In nonrelativistic theories in 
certain models it is known what happens if interactions are included 
\cite{huang93}. In relativistic theories a detailed study of $\lambda
\phi^4$ theory at finite temperature and density has been given in flat
\cite{bensonbernsteindodelson91,bernsteindodelson91} as well as in curved spacetime 
\cite{kirstentoms95}.

Rather than examine self-interacting fields, or the interactions among 
different quantum fields, a simpler problem is to study what happens
for a quantum field under external conditions, where by external conditions we 
mean for example gravitational or electromagnetic background fields or 
simply boundary conditions imposed on the field. Several different 
situations have already been considered. Let us only mention the analysis
in the presence of boundaries and static gravitational fields 
\cite{toms92,toms93,alta,singh,park,cog,dowken,bytvan,kkthesis} 
and in the presence of (mainly constant) magnetic
fields
\cite{schafroth51,schafroth55,may59,may65,toms95,toms95a,daicicfrankelkowalenko94,daicicfrankelgailiskowalenko94,elmforsperssonskagerstam93,elmforsliljenbergperssonskagerstam95,daicic,rojas}.

One of the main questions in all these kinds of considerations is, whether
or not Bose-Einstein condensation can occur. In the above literature
rather involved analysis  has been done in order to arrive at the conclusion
whether or not Bose-Einstein condensation can occur. Recently, the authors
gave a very simple criterion to decide whether a system might condense or not
\cite{physlett}. Here, we plan to give more details of the calculations and to
apply the criterion to a large class of examples. This will include the 
examples treated in the above mentioned 
literature as well as some new cases. Furthermore we will show how the 
criterion might be used to decide if a system reveals the Meissner-Ochsenfeld
effect or not.

It is worth emphasizing precisely what is meant by BEC since different definitions are possible. For the free Bose gas in three or more spatial dimensions the specific heat has a non-smooth behaviour at a critical temperature which signals a phase transition. In terms of the quantum field theory the phase transition may be interpreted as symmetry breaking with the scalar field developing a non-zero expectation value at the critical temperature. Associated with this phase transition is a sudden growth in the occupancy of the ground state. In the present paper we will adopt this as our definition of BEC. It is possible to have a build-up of particles in the ground state without a phase transition occurring. In this case there is no unique way to identify a temperature associated with the build-up of particles in the ground state. For the particular case of a constant magnetic field in three spatial dimensions it was shown by Rojas \cite{rojas} that although there was no phase transition there could still be a build-up of particles in the ground state which could be interpreted as BEC. A similar situation occurs for bosons confined by a harmonic oscillator potential \cite{KirstenTomsPRA}.

\section{Criterion for the occurrence of Bose-Ein\-stein condensation}
\subsection{Relativistic quantum field theory}
Let us consider a complex relativistic scalar field which may interact with
background electromagnetic or gravitational fields but which is otherwise
free. We restrict our attention to an ultrastatic spacetime 
of the form $\cam ={\mathbb R}\times \Sigma$ with metric
\be
ds^2=dt^2 -g_{ij} (x) dx^idx^j \label{2.1}
\ee
and
with the field
obeying any boundary conditions in the spatial directions.
The action functional for the complex field $\Phi$ will be chosen to be
\bea
S&=&\int dt \intsi \left\{
(D^{\mu} \Phi )^{\dagger} (D_{\mu} \Phi )-m^2 \Phi^{\dagger} \Phi
-U_0 (x)\right.\nn\\
&&\quad\quad\quad\left. -U_1 (x) \Phi^{\dagger} \Phi \right\},\label{2.2}
\eea
$d\sigma_x $ is the invariant volume element on the manifold $\Sigma$.
In order to have a scalar field exclusively under external conditions,
the functions $U_0 (x)$ and $U_1 (x)$ may depend on the background
gravitational or electromagnetic fields, but are independent of the scalar
field. They are also assumed to be independent of the time. The kinetic term is the usual gauge-covariant derivative
\be
D_{\mu} \Phi =\partial _{\mu} \Phi -ie A_{\mu} \Phi .\label{2.3}
\ee
In an ultrastatic spacetime, finite temperature and density are easily
incorporated in the Euclidean time formalism 
\cite{bernard74,kapusta89,landsmanvanweert87}. It has been shown, that the 
partition function $\caz$ may be represented in the form 
(see for example \cite{toms94}) 
\be
\caz =\int [d\Phi ][d\Phi^{\dagger}] \exp \left\{-\tilde S\right\},
\label{2.4}
\ee
where
\bea
\tilde S &=& \int\limits_0^{\beta}d \tau \intsi 
\Big\{\left[\dot{\Phi}^{\dagger} +ie(A_0-i\mu)\Phi^{\dagger}\right]\nn\\
&&\quad\quad\times\left[\dot{\Phi}-ie(A_0-i\mu)\Phi\right]
+|{\mathbf D} \Phi |^2\nn\\
&& +(m^2+U_1 (x) ) \Phi^{\dagger} \Phi +U_0 (x)
+J\Phi +J^{\dagger} \Phi^{\dagger}\Big\}.\label{2.5}
\eea
In order to decide whether or not Bose-Einstein condensation will occur we need
the charge in the excited states. It may be obtained directly from the effective
action. Eliminating the dependence on the sources $J$ and $J^{\dagger}$ by
\bea
\back &=& \frac{\delta W}{\delta J}\left.\right|_{\mu,J^{\dagger}}\nn\\
\coba &=& \frac{\delta W}{\delta J^{\dagger}}
\left.\right|_{\mu,J},\label{2.6}
\eea
with the background fields $\back$, $\coba$, the effective action is
defined by
\bea
\Gamma [\mu,\back ,\coba ]& =&W [\mu , J, J^{\dagger}]\nn\\
&& -
\int\limits_0^{\beta} d\tau \intsi [J\back +J^{\dagger} \coba ]
\label{2.7}
\eea
with
\be
W[\mu , J, J^{\dagger}] =-\ln \caz [\mu , J, J^{\dagger}].\label{2.8}
\ee
It is known, that a minimization of the effective action is equivalent to
a minimization of the Helmholtz free energy \cite{toms95}. 
In terms
of the effective action the charge is then defined by
\be
Q=-\frac 1 {\beta} \frac{\partial\Gamma }{\partial \mu} \left.
\right|_{\back, \coba }.\label{2.9}
\ee

We may now proceed with the analysis of the effective action. It is 
straightforward to show that
\be
\Gamma [\mu, \back, \coba ] =\tilde S [\back, \coba ] +\frac 1 2 \ln \det
(l^2 \tilde S _{,ij} ),\label{2.10}
\ee
where we used DeWitt's condensed notation \cite{dewitt65}. $l$ is 
an arbitrary unit of length introduced to keep the argument of the logarithm in (\ref{2.10}) dimensionless. Specializing to the case of a background static magnetic field only, and
choosing the gauge 
\be
A_0 =0, \qquad {\mathbf\nabla}\cdot{\mathbf A }=0,\label{2.11}
\ee
we have
\be
\frac 1 2 \ln\det (l^2 S_{,ij}) =\Gamma_++\Gamma_-,\label{2.11a}
\ee
where 
\bea
\Gamma_{\pm} &=&\frac 1 2 \ln\det\left\{l^2\left[
-\left(\frac{\partial}{\partial \tau} \mp e\mu\right)^2
- {\mathbf D} ^2\right.\right.\nn\\
&&\quad\quad\quad\left.\left. +m^2 +U_1 (x) \right]\right\}.\label{2.12}
\eea
For the calculation of (\ref{2.12}) we will use the zeta function scheme
\cite{hawking77,critchleydowker76}.
In this scheme one defines
\be
\Gamma_{\pm} =-\frac 1 2 \zeta_{\pm} ' (0) +\frac 1 2 \zeta_{\pm} (0) 
\ln l^2 \label{2.13}
\ee
with the  
generalized zeta functions
\be
\zeta_{\pm} (s) =\sum_{j=-\infty}^{\infty}\sum_N (\lambda_{jN} ^{\pm} )^{-s}
.\label{2.14}
\ee
Here, $\lambda _{jN}^{\pm}$ are the eigenvalues of the fluctuation 
operator (see (\ref{2.12})) given by
\be
\lambda_{jN}^{\pm} =\left(\frac{2\pi }{\beta} j \pm ie\mu\right)^2
+\sigma_N,\label{2.15}
\ee
with the eigenvalues $\sigma_N$ of the spatial part of the Klein-Gordon
operator,
\be
(-{\mathbf D}^2 +U_1 (x)+m^2 )f_N (x) =\sigma_N f_N (x) ,\label{2.16}
\ee
with a complete orthonormal set of functions $f_N (x)$. It is easily seen that
$\Gamma_+=\Gamma_-$, so that we skip the index $\pm$ in the notation and
consider $\zeta_+$.

Now all the definitions and preparation to explain our criterion for the
appearance of Bose-Einstein condensation have been given. For the moment we do not
want to consider any specific situation but want to assume a quite
general structure of the eigenvalues $\lambda_{jN}$ or, more specifically,
of the energy $E_N^2 =\sigma_N $. As we will see, to decide if
Bose-Einstein condensation appears, the only relevant information on $E_N$ 
is about its continuous part. Let us assume that $\sigma_N$ splits into the
sum of a discrete part $\sigma_{\mathbf p}^d$ (${\mathbf p}$ is just a set of labels for the 
discrete part of the spectrum), and a continuous part which we can deal with by imposing box normalization. The box will be taken to have sides $L_1,...,L_q$,
for some $q$ and thus we write
\be
E_N^2 =\sum_{i=1}^q \left(\frac{2\pi}{L_i}\right)^2 l_i^2 +\sigma_{\mathbf p}^d\;.\label{2.17}
\ee
In the limit $L_i\to\infty$ the starting point for the analysis of $\zeta (s)$
is
\bea
\zeta (s )&=&\frac{V_q\beta}{\qpi}
\frac 1 {\Gamma (s)} \sul \sum_{\mathbf p} 
\tint t^{s-1-\frac q 2}\nn\\
&&\times \exp\left\{\left(\frac{2\pi il}{\beta}
 -\mu\right)^2
            t -\sigma_{\mathbf p}^d t\right\}.\label{2.18}
\eea
Doing a resummation in $l$, this is equivalent to
\bea
\zeta (s )&=&\frac{V_q\beta}{\qpi}
\frac{\Gamma\left( s-\frac{q+1} 2\right)}
  {\Gamma (s)} \sul \sum_{\mathbf p} \nn\\
& &\times \tint t^{s-1-\frac{ q+1} 2} e^{-\frac{\beta^2}{4t} l^2 
-\sigma_{\mathbf p}^d t +\beta\mu l}\nn\\
&=&\frac{V_q\beta}{\qpi}\frac{\Gamma\left( s-\frac{q+1} 2 \right)}
{\Gamma (s)}\zeta_{d,\Sigma} \left( s-\frac{q+1} 2\right)\nn\\
& &+\frac{V_q\beta}{\qpi} \frac 2 {\Gamma (s)} \sum_{l=1}^{\infty}
\sum_{\mathbf p} \left(\frac 2 {\beta l} \sqrt{\sigma_{\mathbf p}^d}\right)^{\frac{q+1} 2 -s}\nn\\
&&\times K_{\frac{q+1} 2 -s} (\beta l \sqrt{\sigma_{\mathbf p}^d}) \left(e^{\beta\mu l}+e^{-\beta\mu l}\right)\;,\label{2.19}
\eea
where we introduced the zeta-function of the discrete part of the 
spatial section,
\be
\zeta_{d,\Sigma} (s) =\sum_{\mathbf p}(\sigma_{\mathbf p}^d)^{-s}.\label{2.20}
\ee
The condition for Bose-Einstein condensation to occur is that $\mu$ must 
reach a critical value $\mu_C$ set by the lowest eigenvalue in the spectrum,
\be
\mu_C^2 =\sigma_{\mathbf 0}^d =E_0^2  .\label{2.21}
\ee
If the charge $Q$ in the excited states, Eq.~(\ref{2.9}), remains bounded
as $\mu\to\mu_C$, then Bose-Einstein condensation occurs, because for the
total charge large enough, it is not possible to accommodate it all in the 
excited states. If $Q$ is not bounded as $\mu\to\mu_C$, then any amount of
the total charge can reside in the excited states and Bose-Einstein condensation
will not occur. We therefore need to look at the behaviour of $(\partial
/\partial \mu) \zeta (0)$ and $(\partial/\partial\mu )\zeta' (0) $ 
as $\mu\to\mu_C$.

Using the definition (\ref{2.9}) and representation (\ref{2.19}) for
$\zeta (s)$, one immediately finds
\bea
Q&=&V_q\frac 2 {\qpi} \sum_{l=1}^{\infty} \sum_{\mathbf p} \beta l \left(\frac 2 
{\beta l} \sqrt{\sigma_{\mathbf p}^d}\right)^{\frac{q+1} 2} 
K_{\frac{q+1} 2} (\beta l \sqrt{\sigma_{\mathbf p}^d})\nn\\
& &\hspace{2cm}\times \left(e^{l\beta \mu }-e^{-l\beta\mu}\right).\label{2.22}
\eea
The convergence of the sums is defined through the behaviour of the 
MacDonald functions for large arguments \cite{gradshteynryzhik65}
\be
K_{\nu} (z) \sim \sqrt{\frac{\pi}{2z}}e^{-z} \left\{ 1+\cao (z^{-1})
\right\}.\label{2.23}
\ee
It is clear that in the limit $\mu\to\mu_C$ all but the
contributions coming from ${\mathbf p}={\mathbf 0}$ are finite. Thus in the leading approximation,
neglecting finite pieces, we write
\be
Q(\mu\to\mu_C) =d_0V_q\left(\frac{\sqrt{\sigma_{\mathbf 0}^d}}{2\pi\beta}\right)^{q/2}
\sum_{l=1}^{\infty}l^{-q/2} e^{-\beta l (\sqrt{\sigma_{\mathbf 0}^d}-\mu )}.
\label{2.24}
\ee
We have allowed for the possible degeneracy of the ground state by introducing the degeneracy factor $d_0$. It is seen clearly that in the limit $\mu\to\mu_C=\sqrt{
\sigma_{\mathbf 0}^d}$ the charge remains 
finite for $q\geq 3$; thus for $q\geq 3$ Bose-Einstein condensation occurs,
and for $q\leq 2$ Bose-Einstein condensation does not occur.

The detailed behaviour for $\mu\to\mu_C$ is most clearly extracted using the
Mellin-Barnes integral representation of the exponential in (\ref{2.24}),
\be
e^{-v} =\frac 1 {2\pi i} \int\limits_{c-i\infty}^{c+i\infty} d\alpha
\,\,\Gamma(\alpha ) v^{-\alpha},\label{2.25}
\ee
with $\Re v >0$ and $c\in {\mathbb R}$, $c>0$. Using this in Eq.~(\ref{2.24}),
one finds
\bea
Q(\mu\to\mu_C) &=&\frac{d_0V_q}{2\pi i} \left(
\frac{\sqrt{\sigma_{\mathbf 0}^d}}{2\pi\beta}\right)^{\frac q 2}
\int\limits_{c-i\infty}^{c+i\infty}d\alpha\,\,
\Gamma (\alpha )\nn\\
&&\times (\sqrt{\sigma_{\mathbf 0}^d}-\mu )^{-\alpha}\beta^{-\alpha}
\zeta_R \left(\alpha +\frac q 2\right)\;,\label{2.26}
\eea
where, in order to allow for interchanging the sum and the integral one has to
impose $\alpha > 1-q/2$. $\zeta_R(s)$ is the Riemann $\zeta$-function, which is analytic in $s$ except at $s=1$ where it has a simple pole with residue 1. Closing the contour to the left of the rightmost pole
then gives the following leading behaviour~:\\
\\
$q=0$, pole of order one at $\alpha =1$,
\be
Q(\mu\to\mu_C ) =\frac{d_0}{\beta (\mu_C-\mu)},\label{2.27}
\ee
$q=1$, pole of order one at $\alpha =1/2$,
\be
Q(\mu\to\mu_C) =\frac{d_0V_1}{\sqrt{2} \beta} \left(\frac{\mu_C}{\mu_C-\mu}\right)
^{1/2}, \label{2.28}
\ee
$q=2$, pole of order two at $\alpha =0$,
\be
Q(\mu\to\mu_C) =-\frac{d_0V_2}{2\pi\beta} \mu_C\ln\beta (\mu_C-\mu).\label{2.29}
\ee
As mentioned, for $q\geq 3$ no divergent contribution results. The restriction
$q\geq 3$ includes a large number of previously known results, often established
by long and detailed calculations, as special cases. These and some
new examples are summarized in Sec.~3. 

\subsection{Nonrelativistic quantum field theory}

We will consider a nonrelativistic field theory described by the complex Schr\"{o}dinger field $\Phi$ whose action functional is
\bea
S&=&\int dt\int_{\Sigma}d\sigma_x\Big\lbrace\frac{i}{2}\left(\Phi^\dagger \dot{\Phi}-\dot{\Phi}^\dagger\Phi\right)-\frac{1}{2m}|{\mathbf D}\Phi|^2 \nn\\
&&\quad\quad-U_1({\mathbf x})\Phi^\dagger\Phi\Big\rbrace\;.\label{2.2.1}
\eea
$d\sigma_x$ is the invariant volume element on the $D$-dimensional Riemannian manifold $\Sigma$. We will consider $\Sigma$ to be compact. (In the case $\Sigma={\mathbb R}^D$ we will impose box normalization with the infinite box limit taken.) $U_1({\mathbf x})$ is an arbitrary time independent potential. 
As before, 
${\mathbf D}={\mathbf \nabla}-ie{\mathbf A}$ is the gauge covariant derivative, with ${\mathbf A}$ a vector potential describing a background magnetic field
and $|{\mathbf D}\Phi|^2$ is short 
for $g^{ij}D_i\Phi^\dagger D_j\Phi$ where $g_{ij}$ is the Riemannian metric on $\Sigma$.

The theory described by (\ref{2.2.1}) has the local gauge invariance
\bea
\Phi&\rightarrow&e^{ie\theta}\Phi\;,\nn\\
{\mathbf A}&\rightarrow&{\mathbf A}+{\mathbf\nabla}\theta\;,\nn\\
\eea
which gives rise to a conserved current, and a conserved charge which is given by
\be
Q=\int_{\Sigma}d\sigma_x|\Phi|^2\;.\label{2.2.2}
\ee
We will deal with the charge rather than the particle number to mirror the relativistic case considered above as closely as possible. The conserved charge is dealt with by introducing a Lagrange multiplier $\mu$, which is the chemical potential. After a rotation to imaginary time $\tau$, the partition function is again expressed in the form (\ref{2.4}), with
\bea
\tilde{S}\lbrack\bar{\Phi},\bar{\Phi}^\dagger\rbrack&=&\int_{0}^{\beta}d\tau\int_{\Sigma}d\sigma_x\Bigg\lbrace\frac{1}{2}\left(\Phi^\dagger\frac{\partial}{\partial\tau}\Phi-\frac{\partial}{\partial\tau}\Phi^\dagger\Phi\right) \nn\\
&&\quad\quad+\frac{1}{2m}|{\mathbf D}\Phi|^2+U_1({\mathbf x})|\Phi|^2\nn\\
&&\quad\quad-e\mu|\Phi|^2+J\Phi+J^\dagger\Phi^\dagger\Bigg\rbrace\;.
\label{2.2.3}
\eea
The field $\Phi$ has been coupled to a complex source $J$ as in (\ref{2.5}). The effective action can now be defined as in Sec.~2.1. In place of (\ref{2.10}) we find
\bea
\Gamma\lbrack\mu,\bar{\Phi},\bar{\Phi}^\dagger\rbrack&=&
\tilde{S}\lbrack\bar{\Phi},\bar{\Phi}^\dagger\rbrack
+\ln\det l\Big\lbrack
\frac{\partial}{\partial\tau}-e\mu\nn\\
&&\quad-\frac{1}{2m}{\mathbf D}^2+
U_1({\mathbf x})\Big\rbrack\;,\label{2.2.4}
\eea
where $\bar{\Phi}$ is the background Schr\"{o}dinger field. The second term in (\ref{2.2.4}) has arisen from performing the functional integral in (\ref{2.4}) over the Schr\"{o}dinger field. 

In order to regularize the determinant which appears in (\ref{2.2.4}) we will again use the $\zeta$-function method. This time we will let $f_N({\mathbf x})$ denote the eigenfunctions of $\displaystyle{-\frac{1}{2m}{\mathbf D}^2+U_1({\mathbf x})}$~:
\be
\left\lbrack-\frac{1}{2m}{\mathbf D}^2+U_1({\mathbf x})\right\rbrack f_N({\mathbf x})=\sigma_Nf_N({\mathbf x})\;.\label{2.2.5}
\ee
(These $f_N({\mathbf x})$ differ only by a trivial rescaling from the $f_N({\mathbf x})$ used in Sec.~2.1.) The $f_N({\mathbf x})$ are seen to be stationary state solutions to the Schr\"{o}dinger equation for whatever boundary conditions are imposed. The eigenvalues $\sigma_N$ are the energy levels for the first quantized system. The set $\lbrace f_N({\mathbf x})\rbrace$ is assumed to be complete and orthonormal.

Because the functional integral in (\ref{2.4}) extends over all fields periodic in the imaginary time coordinate $\tau$ with period $\beta=T^{-1}$, the eigenvalues of the operator appearing in (\ref{2.2.4}) are
\be
\lambda_{jN}=2\pi ijT+\sigma_N-e\mu\;,\label{2.2.6}
\ee
where $j=0,\pm1,\pm2,\ldots\;$. The generalized $\zeta$-function is given by (see \ref{2.14})
\be
\zeta(s)=\sum_{j=-\infty}^{\infty}\sum_N(\lambda_{jN})^{-s}\;,\label{2.2.7}
\ee
and we define
\FL
\[
\ln\det l\left\lbrack
\frac{\partial}{\partial\tau}-e\mu-\frac{1}{2m}{\mathbf D}^2+
U_1({\mathbf x})\right\rbrack
\]
\FR
\be=-\zeta'(0)+\zeta(0)\ln l\;.\label{2.2.8}
\ee
Using the summation formula in \cite{TomsPRB}, it is easy to show that
\be
\zeta(s)=\sum_N(\sigma_N-e\mu)^{-s}+\zeta_T(s)\;,\label{2.2.9}
\ee
where we have defined
\be
\zeta_T(s)=\frac{T^{-s}}{\Gamma(s)}\sum_N\sum_{n=1}^{\infty}
\frac{e^{-n\beta(\sigma_N-e\mu)}}{n^{1-s}}\;.\label{2.2.10}
\ee
The first term in (\ref{2.2.9}) has no explicit temperature dependence, and corresponds to the zero-point energy contribution to the effective action. We will ignore it in what follows. It disappears if normal ordering of the Hamiltonian is adopted.

The term in $\zeta_T(s)$, which we may call the thermal $\zeta$-function, is easily shown to have the following values~:
\bea
\zeta_T(0)&=&0\;,\label{2.2.11}\\
\zeta_T'(0)&=&-\sum_N\ln\left\lbrack1-e^{-\beta(\sigma_N-e\mu)}\right\rbrack\;.\label{2.2.12}
\eea
These last two results show that the second term in (\ref{2.2.4}) has a simple relation to the thermodynamic potential $\Omega$ defined by
\be
\Omega=T\sum_N\ln\left\lbrack1-e^{-\beta(\sigma_N-e\mu)}\right\rbrack\;. \label{2.2.13}
\ee
(We could have adopted $\Omega$ as our starting point; however, our criterion is most simply expressed using the generalized $\zeta$-function, and in addition we wished to parallel the relativistic calculation.)

We will again assume that $\sigma_N$ splits up into the sum of a discrete part $\sigma_{\mathbf p}^d$, and a continuous part which is dealt with by box normalization as in (\ref{2.17}) with $E_N^2$ replaced by $\sigma_N$. In the large box limit, the thermal $\zeta$-function becomes
\be
\zeta_T(s)=\frac{V_q}{(4\pi)^{q/2}}\frac{T^{q/2-s}}{\Gamma(s)}
\sum_{\mathbf p}\sum_{n=1}^{\infty}\frac{e^{-n\beta(\sigma_{\mathbf p}^d - e\mu)}}{n^{1+q/2-s}}\;.\label{2.2.14}
\ee
As in the relativistic case, the lowest mode $\sigma_{\mathbf 0}^d$ plays the crucial role in determining whether or not Bose-Einstein condensation can occur. We therefore split the sum over $\mathbf p$ in (\ref{2.2.14}) by separating off the lowest mode for special treatment. The critical value of the chemical potential is given by
\be
e\mu_C=\sigma_0=\sigma_{\mathbf 0}^d\;,\label{2.2.15}
\ee
and we wish to study the behaviour of the charge as $\mu\rightarrow\mu_C$. If we write
\be
\zeta_T(s)=\zeta_T^{(0)}(s)+\zeta_T^{(\ne0)}(s)\;,\label{2.2.16}
\ee
where
\be
\zeta_T^{(0)}(s)=\frac{d_0V_q}{(4\pi)^{q/2}}\frac{T^{q/2-s}}{\Gamma(s)}
\sum_{n=1}^{\infty}\frac{e^{-n\beta e(\mu_C- \mu)}}{n^{1+q/2-s}}\label{2.2.17}
\ee
is the lowest mode contribution (with degeneracy $d_0$), and $\zeta_T^{(\ne0)}(s)$ is given by (\ref{2.2.14}) with the sum over $\mathbf p$ restricted to non-zero values, it is easy to see that because the argument of the exponential in (\ref{2.2.14}) will always be negative, even for $\mu=\mu_C$, $\zeta_T^{(\ne0)\prime}(0)$ and $\displaystyle{\frac{\partial}{\partial\mu}\zeta_T^{(\ne0)\prime}(0)}$ remain finite. Bose-Einstein condensation is therefore determined from a knowledge of $\zeta_T^{(0)}(s)$.

From (\ref{2.2.17}) we have $\zeta^{(0)}(0)=0$ and
\be
\zeta_T^{(0)\prime}(0)=\frac{d_0V_q}{(4\pi)^{q/2}}T^{q/2-s}
\sum_{n=1}^{\infty}\frac{e^{-n\beta e(\mu_C- \mu)}}{n^{1+q/2}}
\;.\label{2.2.18}
\ee
The charge is given by
\be
Q=-T\frac{\partial}{\partial\mu}\Gamma=T\frac{\partial}{\partial\mu}
\zeta_T^{(0)\prime}(0)+\cdots\;,\label{2.2.19}
\ee
where terms which remain finite as $\mu\rightarrow\mu_C$ have been dropped. Using (\ref{2.2.18}) we have
\be
Q(\mu\rightarrow\mu_C)\simeq ed_0V_q\left(\frac{T}{4\pi}\right)^{q/2}\sum_{n=1}^{\infty}
\frac{e^{-n\beta e(\mu_C- \mu)}}{n^{q/2}}
\;.\label{2.2.20}
\ee
At $\mu=\mu_C$ the sum appearing in this expression for the total charge diverges for $q\le2$. This gives us the necessary and sufficient condition for Bose-Einstein condensation to occur. For $q\ge3$, Bose-Einstein condensation does not occur.

We can obtain more detailed information on how $Q$ diverges as $\mu\rightarrow\mu_C$ as before by making use of (\ref{2.25}). It is easily seen that (\ref{2.2.20}) becomes
\bea
Q(\mu\rightarrow\mu_C)&=&\frac{ed_0V_q}{2\pi i}\left(\frac{T}{4\pi}\right)^{q/2}
\int_{c-i\infty}^{c+i\infty}d\alpha\Gamma(\alpha)\nn\\
&&\times\left(\frac{T}{\mu_C-\mu}\right)^\alpha\zeta_R(\alpha+q/2)
\;.\label{2.2.21}
\eea
This result may be used to show that
\bea
Q(\mu\rightarrow\mu_C)&\simeq&\frac{eTd_0}{\mu_C-\mu}\quad (q=0);\label{2.2.22}\\
Q(\mu\rightarrow\mu_C)&\simeq&\frac{1}{2}ed_0V_1T(\mu_C-\mu)^{-1/2}\quad (q=1);\label{2.2.23}\\
Q(\mu\rightarrow\mu_C)&\simeq&\frac{eTd_0V_2}{4\pi}\ln\left(\frac{T}{\mu_C-\mu}\right)\quad (q=2).\label{2.2.24}
\eea
Only the leading part of $Q$ which diverges as $\mu\rightarrow\mu_C$ has been shown in these expressions. The way in which $Q$ diverges as $\mu\rightarrow\mu_C$ is seen to be the same for relativistic and nonrelativistic systems.

\section{Application of the criterion to examples}

In this section we will consider several examples of an ideal Bose gas
under possible external conditions and we will see that the conclusion
whether or not Bose-Einstein condensation might occur can be drawn very
easily using our criterion of section 2.

Let us start with the free ideal Bose gas in a $(D+1)$-dimensional Min\-kow\-ski
space-time. Then the eigenvalues of the Klein-Gordon operator (\ref{2.16})
are 
\be
E_{\vec k}^2 =\vec k ^2 +m^2,\qquad \vec k \in {\mathbb R} ^D.\label{4.1}
\ee
In the non-relativistic case we have
\be
E_{\vec k}=\frac{1}{2m}{\vec k}^2,\qquad \vec k \in 
{\mathbb R}^D\;.\label{4.1nr}
\ee
In either case we see that $q=D$, and conclude that BEC can only occur for $D\ge3$. This agrees with conclusions of Refs.~\cite{haberweldon81,haberweldon82a,kapusta81} for the relativistic gas, and Refs.~\cite{may59,Ziff} for the non-relativistic gas.

There are many possibilities of restricting the Minkowski space
by imposing boundary conditions in one or more directions. For example
imposing Dirichlet boundary conditions in one direction, 
$f_N (x_i=0) =f_N (x_i=L)$, the energy 
eigenvalues eq.~(\ref{2.17}) are 
\be
E_{n,\vec k}^2 = \vec k^2 +\left(\frac{\pi n}{L}\right)^2 +m^2,
\,\,\vec k \in{\mathbb R} ^{D-1},\,\, n\in{\mathbb N} ,\label{4.2}
\ee
for the relativistic field, and
\be
E_{n,\vec k}=\frac{1}{2m}\left\lbrace{\vec k}^2+
\left(\frac{\pi n}{L}\right)^2\right\rbrace,\,\,\vec k \in{\mathbb R} ^{D-1},\,\, n\in{\mathbb N} ,\label{4.2nr}
\ee 
for the non-relativistic field. In this case we find $q=D-1$ and therefore BEC is only expected for $D\ge4$ (corresponding to $q\ge3$). If we impose Dirichlet boundary conditions in $p$ of the spatial dimensions then $q=D-p$ and BEC would only be expected for $D\ge3+p$. The choice of Dirichlet boundary conditions is not important here, and periodic or Neumann boundary conditions would lead to the same result. The restriction $D\ge3+p$ holds for any choice of boundary conditions which results in a discrete spectrum in $p$ spatial dimensions.

Another important case where the energy spectrum contains a discrete part is found when there is a constant external magnetic field present. Starting with a single
component constant magnetic field $B$ one encounters the well known Landau
levels giving
\be
E_{n,\vec k _{\bot}}^2
=(2n+1)eB+\vec k _{\bot}^2+m^2,\,\,n\in{\mathbb N}_0,\,\,
\vec k _{\bot} \in{\mathbb R} ^{D-2},\label{4.3}
\ee
for the relativistic field, and
\be
E_{n,\vec k_{\bot}}=\frac{1}{2m}\left\lbrace(2n+1)eB+{\vec k}_{\bot}^2\right\rbrace\,\,\label{4.3nr}
\ee
in the nonrelativistic case. In both cases the eigenvalues are degenerate with degeneracy $(SeB)/(2\pi)$ with the (formal) volume $S$ of 
${\mathbb R} ^2$. The single component magnetic field results in $q=D-2$, and condensation can occur only for $D\geq 5$. This was shown for the relativistic field in \cite{daicicfrankelkowalenko94,daicicfrankelgailiskowalenko94} and for the non-relativistic field in \cite{may65}. In particular BEC is absent for $D=3$ \cite{schafroth55}.

In the general case, a magnetic field in $D$ spatial dimensions is characterized by $\delta$ independent components, where $D=2\delta$ or $2\delta+1$ depending upon whether $D$ is even or odd \cite{toms95a,toms94}. It is straightforward to show that
\be
E_{n,\vec k _{\bot}}^2=\sum_{j=1}^l (2n_j +1) eB_j +\vec k_{\bot}^2+m^2
,\label{4.4}
\ee
for the relativistic field, and
\be
E_{n,\vec k_{\bot}}=\frac{1}{2m}\left\lbrace
\sum_{j=1}^l (2n_j +1) eB_j +\vec k_{\bot}^2\right\rbrace
,\label{4.4nr}
\ee
for the non-relativistic field. In both cases $n_j\in{\mathbb Z}_0$ and $\vec k _{\bot} \in{\mathbb R} ^{D-2l}$. The degeneracy is $\prod_{j=1}^{p}eB_jL_{2j-1}L_{2j}/(2\pi)$ where $B_j\ j=1,\ldots,p$ represent the independent field strengths and $p\le\delta$. ($L_j$ is the length of the box in the $j^{\rm th}$ direction with $L_j\rightarrow\infty$ assumed.) In either case, $q=D-2p$, so that every component of the magnetic field reduces the effective dimension by 2. The criterion for BEC is $D\ge3+2p$ in agreement with the explicit calculations in Refs.~\cite{toms95a,toms94}.

Another example is to consider a field on the space ${\mathbb S}^{p}\times{\mathbb R}^{D-p}$ where ${\mathbb S}^{p}$ is the $p$-dimensional sphere. In this case we have
\be
E_{n,\vec k}^2=n(n+p-1)a^{-2}+{\vec k}^2+\xi a^{-2}+m^2\,\label{einrel}
\ee
for the relativistic field, and
\be
E_{n,\vec k}=\frac{1}{2m}\left\lbrack n(n+p-1)a^{-2}+\xi a^{-2}+\vec k^2\right\rbrack\;,\label{einnr}
\ee
for the non-relativistic field. $a$ represents the radius of the sphere, and $\xi$ is a dimensionless coupling constant which is related to a possible coupling of the scalar field to the scalar curvature. We have $q=D-p$ here, and BEC requires $D\ge3+p$. In the case of greatest physical interest we have $p=D=3$ which corresponds to the Einstein static universe. BEC does not occur in this case as shown recently in Ref.~\cite{smithtoms}.

In the examples we have discussed so far the energy spectrum has been explicitly known so that a direct evaluation of the thermodynamics has been possible. Our next example is to consider a case where the spectrum is not explicitly 
known; namely a single component constant magnetic field in a cylindrical 
box of radius $R$, where the field satisfies Dirichlet boundary
conditions \cite{toms94}.
The eigenvalues for this problem are given in implicit form only and the 
equation reads
\be
M\left( \frac n 2 -\frac 1 4 \gamma , n+1; \frac 1 2 eBR^2\right) =0 
\label{4.5}
\ee
with $n\in{\mathbb N}$, $\gamma =(2/eB)[E_N-eB(n+1)-\vec k _{\bot}^2 ]$,
$\vec k _{\bot}^2 \in{\mathbb R} ^{D-2}$,
and $M(a,b;c)$ the confluent hypergeometric function 
\cite{gradshteynryzhik65}. The zeroes of (\ref{4.5}) are a discrete 
set of values for $\gamma$; thus $q=D-2$ and condensation occurs 
for $D\geq 5$, a criterion derived without any problem. The exact analysis of the problem would be impossible due to the lack of an explicit form for the zeros of the confluent hypergeometric function.

A similar situation to the last example occurs if we consider a gas confined by an arbitrarily shaped cavity of dimension $p$ 
in a space of dimension $D$. It is impossible to write down  any explicit form for the energy eigenvalues. If the confinement is achieved by imposing Dirichlet boundary conditions say, then the portion of the energy spectrum coming from the cavity would be discrete and $q=D-p$. A more complete analysis of this problem will be given elsewhere \cite{KKDJTinprep}.

As our final example we consider a non-relativistic gas in a harmonic oscillator potential. If we assume that
\be
U({\mathbf x})=\frac{1}{2}m\sum_{j=1}^{p}\omega_j^2x_j^2\;,\label{pot}
\ee
with $p\le D$, then
\be
E_{n,\vec k}=\frac{1}{2m}\left\lbrace\vec k^2+\sum_{j=1}^{p}(n_j+\frac{1}{2})\omega_j\right\rbrace\;,\ \ \vec k\in{\mathbb R}^{D-p}\;,\label{spect}
\ee
are the energy levels. We have $q=D-p$ here so that $D\ge3+p$ is 
required for BEC. In the case of greatest physical interest $D=3$ 
it can be seen that BEC does not occur. At this stage it is worth repeating the comment in the last paragraph of Sec.~1. We are interpreting BEC in the sense of symmetry breaking with an associated phase transition, such as that which occurs for the free Bose gas with $D=3$. This does not occur unless $D\ge3+p$. We are not saying that there cannot be a build-up of particles in the 
ground state unassociated with a phase transition as the temperature is lowered.

This concludes the list of examples for the relativistic and non-relativistic Bose gas. As we 
have seen, known examples and in addition new examples can be dealt
with very easily.

\section{Appearance of the Meissner-Ochsenfeld effect for a general ho\-mo\-ge\-ne\-ous
mag\-net\-ic field}
\subsection{Relativistic scalar field in an external magnetic field}
Let us consider the situation where a general homogeneous magnetic field is 
present. The energy eigenvalues then read 
\be
E_{N}^2 =\sum_{j=1}^p (2n_j+1) \nu_j +\sum_{j=2p+1}^D 
\left(\frac{2\pi l_j}{L_j}\right)^2 +m^2 \label{3.1}
\ee
with the limit $L_j =L \to \infty$ taken and $\nu_j =eB_j$. $p$ is the number of non-zero components of the magnetic field as discussed in the last section. The degeneracy 
for each Landau level is $(L^2/(2\pi ))^p\prod_{j=1}^p \nu_j $. (This degeneracy corresponds to $d_0$ introduced in Sec.~2.) Thus for the
zeta function we have
\bea
\zeta (s) &=& \frac{L^D}{2^{D-p}\pi^{D/2}\Gamma (s)}\prod_{j=1}^p \nu_j
\sul \snenp \times\label{3.2}\\
& &\int\limits_0^{\infty} dt\,\, t^{s-1+p-D/2} e^{\left(\frac{2\pi il}{\beta}
-\mu\right)^2 t -\sum_{j=1}^p (2n_j +1) \nu_j t -m^2 t}.\nn
\eea
For the magnetization
\be
M_i =-\frac 1 {L^D\beta} \frac{\partial}{\partial B_i } \Gamma \label{3.3}
\ee
we need $(\partial /\partial B_i) \zeta (s)$. (The magnetization and the choice of units is described in the Appendix.)
Using the same kind of steps as for the derivation of 
(\ref{2.19}), we find
\bea
\frac{\partial}{\partial B_i} \zeta (s)
&=&\frac{L^D\beta}{2^{D+1-p}\pi^{\frac{D+1} 2} \Gamma (s)}
\left(\pnenj\right)\nn\\
&&\quad\times\left\{ -e \Gamma
\left(s-\frac{q-1} 2 \right)\tilde{\zeta} _{d,\Sigma}\left( 
\frac{q-1} 2 -s\right)\right.\nn\\
& &\quad\left. +\frac 1 {B_i} \Gamma\left( s-\frac{q+1} 2\right)
\zeta_{d,\Sigma} \left(\frac{q+1} 2 -s\right)\right\}\nn\\
& &+\frac{L^D\beta}{2^{\frac{D+1} 2 +s} \pi^{\frac{D+1} 2}\Gamma (s)} 
\pnenj \sum_{l=1} ^{\infty} \snenp e^{\beta \mu l}\nn\\
&&\quad\times \left(\frac{\sqrt{\sigma_{\mathbf p}^d}}{\beta l}\right)^{\frac{q-1} 2 -s}\Bigg\{-e(2n_i+1)\nn\\
&&\quad\quad\times K_     
{\frac{q-1} 2 -s } (\beta l \sqrt{\sigma_{\mathbf p}^d})\nn\\
&&\quad+\frac 2 {B_i} \left(\frac{\sqrt{\sigma_p^d}}{\beta l}\right) 
 K_{\frac{q+ 1} 2 -s } (\beta l \sqrt{\sigma_{\mathbf p}^d})\Bigg\}     
\nn\\
& &+(\mu\to-\mu),\label{3.4}
\eea
where we defined
\be
\tilde{\zeta} _{d,\Sigma} (s) =\snenp (2n_i+1)(\sigma_{\mathbf p}^d)^{-s},\label{3.5}
\ee
and abbreviated to $(\mu\to-\mu)$ the terms obtained from replacing $\mu$ in the second set of terms in (\ref{3.4}) with $-\mu$. In order to decide whether or not the Meissner-Ochsenfeld effect takes place we can consider the limit $B_i \to 0$ and see if a nonvanishing
magnetization results. This is equivalent to the generalization of the 
Meissner-Ochsenfeld effect from $D=3$ dimensions 
\cite{elmforsliljenbergperssonskagerstam95} to arbitrary dimension $D$. When $B_i\rightarrow0$ this removes one of the discrete degenerate Landau levels and replaces it with two additional continuous labels. (See eq.~(4.1).) The effect of this is to increase $q$ by 2 to $q_{eff}=q+2$. Our general analysis showed that the gas could only condense for $q\ge3$. Thus if $q_{eff}\ge3$, or $q=1,2$ there may be interesting behaviour when a magnetic field is present even though strict BEC cannot occur. The case $q=1$ in the limit $B_i\rightarrow0$ can be compared  with the gas for $B_i=0$ with $q=3$; the case $q=2$ in the limit $B_i\rightarrow0$ can be compared with the gas for $B_i=0$ with $q=4$. In either case we wish to study the behaviour of the magnetization as $\mu$ becomes close to the critical value $\mu_c$. As argued in our criterion for BEC, the leading term comes from the smallest energy eigenvalue and it reads
\bea
M_i (B_i \to 0)&=&\frac{e\beta^{-q/2}}{2^{\frac{D+2} 2 } \pi^{D/2}}
\left(\pnenj \right) \mu ^{\frac{q-2} 2}\nn\\
&&\quad\times \sum_{l=1}^{\infty} l^{-q/2} 
e^{-\beta l (\mu_C-\mu)},\label{3.6}
\eea
which is very similar to the behaviour of the charge, Eq.~(\ref{2.24}).
We see that for $q\geq 3$ the sum is convergent, even for $\mu =\mu_C$, and
thus $M_i (B_i \to 0) =0$ for that case. However, for $q=1,2,$ we find
using once more (\ref{2.25}),\\
\\
$q=1$
\be
M_i (B_i \to 0) =\frac{eT}{2^{\frac{D+2} 2 } \pi^{\frac{D-1} 2}}
\left(\pnenj\right)\frac 1 {\sqrt{\mu_C} \sqrt{\mu_C-\mu}},\label{3.7}
\ee
$q=2$
\be
M_i (B_i \to 0) =-\frac{eT}{2^{\frac{D+2} 2 } \pi^{\frac{D} 2}}
\left(\pnenj\right)\ln (\mu_C-\mu ).\label{3.8}
\ee
For $q=0$ in the limit $B_i \to 0$ the effective dimension gets increased to $q=2$ and
thus no condensation can occur. Thus $\mu < \mu_C$, the sums are convergent
and though $M_i (B_i \to 0) =0$ in that case in the given approximation.

The magnetization laws, Eqs.~(\ref{3.7}) and (\ref{3.8}), strongly 
depend on how $\mu\to\mu_C$ for $B_i\to 0$. It is clear, that the 
charge in the continuum above the lowest Landau level will condense into
the ground state in the limit $B_i \to 0$. Let us use the notation
$Q_{gr}(\mu\to\mu_C)$ for that charge. This is however exactly the charge
given in Eqs.~(\ref{2.28}) and (\ref{2.29}), and the limit
$M_i (B_i \to 0)/Q_{gr} (\mu\to\mu_C)$ is seen to be finite. If we introduce the charge density $\rho_{gr}=Q_{gr}/L^D$, it reads for $q=1,2$, 
\be
\frac{M_i (B_i \to 0)}{\rho_{gr} (\mu\to\mu_C)} =-\frac e {2\mu_C}.\label{3.9}
\ee
Using the universal knowledge of the charge $Q_{gr}$ in the ground state 
as a function of the critical temperature \cite{toms92,toms93},
\be
Q_{gr} =Q \left(1-\left(\frac T {T_c}\right)^ {D-1} \right)\label{3.10}
\ee
with the total charge $Q$, and total charge density $\rho=Q/L^D$, we arrive at
\be
M_i (B_i \to 0) =-\frac e {2\mu_C} \rho 
\left(1-\left(\frac T {T_c}\right)^ {D-1} \right),             
\label{3.11}
\ee
with $\mu_C =\sqrt{m^2+\sum_{j=1,j\neq i}^p \nu_j}$ as the magnetization law
for $q=1,2$.
For $q=1$ in three spatial dimensions ($D=3$) the result reduces to the one of
ref.~\cite{elmforsliljenbergperssonskagerstam95}, and we agree with their
conclusions on the Meissner-Ochsenfeld effect.
       
\subsection{Non-relativistic scalar field}

The energy levels for a $p$-component constant magnetic field are
\be
E_{n,l}=\sum_{j=1}^{p}(2n_j+1)\frac{eB_j}{2m}+\frac{1}{2m}
\sum_{j=2p+1}^{D}\left(\frac{2\pi l_j}{L_j}\right)^2\;,
\ee
where $n_j=0,1,\ldots$, and $L_j=L\rightarrow\infty$ as before. The degeneracy is $L^{2p}\prod_{j=1}^{p}eB_j/(2\pi)$. The effective action is determined as described in Sec.~2.2, and the magnetization found as in (4.3). We can again argue that $q=1,2$ are the interesting values as $B_i\rightarrow0$ since in this limit we obtain $q_{eff}=q+2=3,4$ for which BEC is possible. 
The lowest state contribution is of greatest interest in this limit and we find
\bea
M_i(B_i\rightarrow0)&\simeq&-\frac{e}{2m}
\left(\frac{m}{2\pi\beta}\right)^{q/2}
\left(\frac{eB_j}{2\pi}\right)\nn\\
&&\quad\times\sum_{n=1}^{\infty}
\frac{e^{-n\beta e(\mu_c-\mu)}}{n^{q/2}}\;,\label{star}
\eea
as the leading contribution. (Note that $\mu_c=\frac{1}{2m}\sum_{j=1}^{p}eB_j$ here.) Apart from a numerical prefactor, this is the same as in the relativistic case in (4.6). For $q\ge3$ the sum converges and we find $M_i\rightarrow0$ as $B_i\rightarrow0$ even if $\mu=\mu_c$. The sum occurring in (\ref{star}) is the same as that occurring in the ground state charge (2.51).. If we write $\rho_{gr}=Q(\mu\simeq\mu_c)/L^D$ as the ground state charge density we find
\be
M_i(B_i\rightarrow0)\simeq-\frac{\rho_{gr}}{2m}\;.
\ee
This is essentially the same as the result found by Schafroth \cite{schafroth55} for $D=3$.

\section{Conclusions}

In this article we considered the general setting of a quantum field, 
relativistic or nonrelativistic, without self-interactions under general
external conditions. We tried to answer the question whether or not the system
might undergo a Bose-Einstein condensation, where by Bose-Einstein condensation we mean a phase transition associated with a build-up of particles in the ground state. Our main result is that the crucial feature governing Bose-Einstein condensation is the dimension
$q$ associated with the continuous part of the eigenvalue spectrum of the 
Hamiltonian for nonrelativistic systems or the spatial part of the Klein-Gordon
operator for relativistic systems. In either case Bose-Einstein condensation
can only occur if $q\geq 3$. Having this criterion at hand, many of previous results
were obtained easily in section 3. In addition some new applications were given. The criterion was shown by studying the lowest eigenvalue of a differential
operator. This relates to the idea of the effective infrared dimensions 
and finite size effects studied earlier \cite{Hu}.

Furthermore, we applied our ideas to the appearance of the Meissner-Ochsenfeld
effect for a general homogeneous magnetic field. With the help of our 
criterion a discussion of this effect is greatly simplified and special
cases are easily recovered 
\cite{schafroth55,elmforsliljenbergperssonskagerstam95,daicic}.

The extension of our method to study Bose-Einstein condensation in theories with
self-interaction 
\cite{bensonbernsteindodelson91,bernsteindodelson91} is of obvious interest. 
Recently we have shown how this
problem can be tackled very efficiently using $\zeta$-function methods 
\cite{kirstentoms95}. It is very likely that the method described in the present paper 
can be extended to interacting field theory. In particular it is possible
to define an effective field theory describing the lowest modes 
as in ref. \cite{Hu}. 

\section*{Acknowledgments}
The work of KK is supported by the DFG under contract number Bo 1112/4-1.

\appendix
\section*{A note about units}

\renewcommand{\H}{\mathbf H}
\newcommand{\B}{\mathbf B}
\newcommand{\M}{\mathbf M}
\newcommand{\J}{\mathbf J}

As in classical electromagnetism, the magnetization $\M$ causes an effective current density which changes the effective field strength. This was noted by Schafroth \cite{schafroth55} in the non-relativistic case, who argued that because the radius of the classical orbit for the charged particles was much greater than the average interparticle separation (whose scale is set by the particle density), the effective field strength should be identified with the average microscopic field. In the opposite case, where the interparticle separation is much greater than the classical charged particle orbit, then the correct procedure is to treat the particles as point dipoles. 

An important point is that Schafroth used cgs units for his electromagnetic field conventions. In this case (see Jackson \cite{Jackson}) the magnetization gives rise to an effective current density ${\J}_M=c\nabla\times{\M}$ which must be added to the current in the Maxwell equation $\nabla\times{\B}={\displaystyle\frac{4\pi}{c}}{\J}$. The resulting Maxwell equation may be written as $\nabla\times{\H}={\displaystyle\frac{4\pi}{c}}{\J}$ where
\begin{equation}
\H=\B-4\pi\M\;.\label{eq:B1}
\end{equation}
An important point is that the factor of $4\pi$ which enters here has nothing to do with the fact that we are in three spatial dimensions. It is simply a consequence of the fact that cgs units have been chosen. In SI units we make the changes \cite{Jackson} 
\begin{eqnarray*}
\B&\rightarrow&\sqrt{\frac{4\pi}{\mu_0}}\ \B\\
\M&\rightarrow&\sqrt{\frac{\mu_0}{4\pi}}\ \M\\
\H&\rightarrow&\sqrt{4\pi\mu_0}\ \H
\end{eqnarray*}
and (\ref{eq:B1}) is changed to
\begin{equation}
\H=\frac{1}{\mu_0}\B-\M\;.\label{eq:B2}
\end{equation}
In this case there is no factor of $4\pi$ in front of $\M$.

Our reason for making such a fuss over the units is that Ref.~\cite{daicicfrankelkowalenko94,daicicfrankelgailiskowalenko94} has altered (\ref{eq:B1}) to read
\begin{equation}
\H=\B-\frac{D\pi^{D/2}}{\Gamma(1+D/2)}\M\label{eq:B3}
\end{equation}
in $D$ spatial dimensions. This expression first appears in the work of May \cite{may65} who generalized the original calculation of Schafroth \cite{schafroth55} from 3 to $D$ dimensions. The $D$-dependent factor in (\ref{eq:B3}) was noted as the area of the $(D-1)$-dimensional sphere by May. However, it should be clear that the factor appearing in front of the magnetization $\M$ is due to the choice of units, nothing more. (For the purist, we note that all of the above can be rewritten in terms of differential forms with the exterior derivative ${\mathbf d}$ in place of $\nabla\times$.)

There is another simple way of deducing the effective magnetic field which bears out our interpretation. The energy stored in the magnetic field is
\begin{equation}
W=\frac{1}{8\pi}\int_{\Sigma}d\sigma_x\,\B\cdot\B\label{eq:B4}
\end{equation}
in cgs units \cite{Jackson}. If $\B$ is constant, then we have simply
\begin{equation}
W=\frac{1}{8\pi}VB^2\;,\label{eq:B5}
\end{equation}
with $B=|\B|$. This term should really be added to the free energy $F$ computed from the partition function or thermodynamic potential to form
\begin{equation}
F_T=F+\frac{1}{8\pi}VB^2\;.\label{eq:B6}
\end{equation}
The magnetization is identified by the usual thermodynamic expression \cite{Gug} to be
\begin{equation}
M=-\frac{1}{V}\frac{\partial F}{\partial B}\;,\label{eq:B7}
\end{equation}
and therefore we have
\begin{equation}
\frac{\partial F_T}{\partial B}=\frac{V}{8\pi}(B-4\pi M)\;.\label{eq:B8}
\end{equation}
In our calculation, $M=|\M|$ has arisen from one-loop quantum effects. In the absence of quantum effects we would have simply the term in (\ref{eq:B8}) which involves $B$. This suggests that we define the effective field strength as in (\ref{eq:B1}). If SI units are adopted the factors of $\displaystyle{\frac{1}{8\pi}}$ in (\ref{eq:B4}--\ref{eq:B6}) are replaced with $\displaystyle{\frac{1}{2\mu_0}}$, and a repeat of this simple argument identifies (\ref{eq:B2}) as the effective field strength.

In relativistic quantum field theory it is conventional to use neither cgs nor SI units, but instead Heaviside-Lorentz rationalized units. In these units, where the lagrangian density for electromagnetism is $\displaystyle{-\frac{1}{4}F_{\mu\nu}F^{\mu\nu}}$, (\ref{eq:B6}) becomes
\begin{equation}
F_T=F+\frac{1}{2}VB^2\;,\label{eq:B9}
\end{equation}
and we identify
\begin{equation}
H=B-M\label{eq:B10}
\end{equation}
as the effective field strength. (The difference between Heaviside-Lorentz rationalized units and cgs units is that $B_{\rm cgs}=\sqrt{4\pi}\,B_{\rm HL}$, with a similar scaling relation for the vector gauge fields $A_\mu$. See Schweber \cite{Schweber} for example.) We adopt Heaviside-Lorentz rationalized units in our work, as usual in quantum field theory. 

We will now show how the effective action formalism may be used to calculate the magnetization for a general applied magnetic field in a space of arbitrary dimension. The full effective action may be written as
\begin{equation}
\Gamma=S_{\rm em}+\tilde{S}+\tilde{\Gamma}\;\label{eq:B11}
\end{equation}
where
\begin{equation}
S_{\rm em}=\beta\int_{\Sigma}d\sigma_x\left\lbrace\frac{1}{4}F_{ij}F^{ij}- 
J_{\rm ext}^iA_i\right\rbrace\label{eq:B12}
\end{equation}
is the electromagnetic field action in Heaviside-Lorentz rationalized units. $F_{ij}$ is the magnetic field tensor in a Riemannian space $\Sigma$ with the time taken to be imaginary with periodicity $\beta$. We assume that the magnetic field is static, but otherwise arbitrary at this stage. $J_{\rm ext}^i$ is the externally applied current responsible for setting up the magnetic field. $\tilde{S}$ represents the contribution from the background scalar field, or Schr\"{o}dinger field. $\tilde{\Gamma}$ represents the contribution from quantum effects of the matter fields.

We can settle the question of the correct expression for the magnetization in $D$ spatial dimensions by working out the Maxwell equations for the magnetic field with quantum effects included. We simply need to evaluate $\delta\Gamma$ when $A_i$ or $F_{ij}$ is varied, and set the variation equal to zero, since this will give the effective field equation. Variation of (\ref{eq:B12}) leads to
\begin{equation}
\delta S_{\rm em}=\beta\int_{\Sigma}d\sigma_x\left\lbrace\nabla_jF^{ij}-J_{\rm ext}^i\right\rbrace\delta A_i\;.\label{eq:B13}
\end{equation}
Here $\nabla_i$ is the covariant derivative computed using the metric on $\Sigma$. We will define the ground state contribution to the current density by
\begin{equation}
\delta\tilde{S}=-\beta\int_{\Sigma}d\sigma_x\,J_{\rm ground}^i\delta A_i\;,
\label{eq:B14}
\end{equation}
as in Ref.~\cite{TomsPRB}. We now need to compute $\delta\tilde{\Gamma}$. In the actual calculation of $\tilde{\Gamma}$, the result is expressed in terms of $F_{ij}$ rather than directly in terms of $A_i$. However, it is easy to prove that
\begin{equation}
\frac{\delta\tilde{\Gamma}}{\delta A_i}=2\nabla_j\left(
\frac{\delta\tilde{\Gamma}}{\delta F_{ij}}\right)\;.\label{eq:B15}
\end{equation}
If we define the current induced by quantum effects $J_{\rm ind}^i$ by \cite{TomsPRB}
\begin{equation}
J_{\rm ind}^i=-\frac{2}{\beta}\nabla_j\left(\frac{\delta\tilde{\Gamma}}
{\delta F_{ij}}\right)\;,\label{eq:B16}
\end{equation}
then the effective Maxwell equation obtained from $\displaystyle{\frac{\delta\Gamma}{\delta A_i}}=0$ is
\begin{equation}
\nabla_jF^{ij}=J_{\rm ext}^i+J_{\rm ground}^i+J_{\rm ind}^i\;.\label{eq:B17}
\end{equation}

It is possible to rewrite (\ref{eq:B17}) in a form which resembles the result in the case of $D=3$. This is done by defining the tensor $H_{ij}$ by
\begin{equation}
H^{ij}=F^{ij}+2T\frac{\delta\tilde{\Gamma}}{\delta F_{ij}}\;.\label{eq:B18}
\end{equation}
As we will show in a moment, this is the analogue of the vector $\H$ in the usual $D=3$ case. With the definition in (\ref{eq:B18}), we can rewrite (\ref{eq:B17}) as
\begin{equation}
\nabla_jH^{ij}=J_{\rm ext}^i+J_{\rm ground}^i\;.\label{eq:B19}
\end{equation}
In order to see that our definitions correspond to the usual ones for $D=3$, described in the earlier part of this appendix, restrict attention to $D=3$. The magnetic field vector $\B$ with components $B^i$ is defined in terms of the field strength tensor $F_{ij}$ by $F_{ij}=\epsilon_{ijk}B^k$, or equivalently by $B^i=\displaystyle{\frac{1}{2}}\epsilon^{ijk}F_{jk}$, where $\epsilon_{ijk}$ is the Levi-Civitta tensor. In the same way we can define the vector $\H$ with components $H^i$ in terms of the tensor $H_{ij}$ by $H_{ij}=\epsilon_{ijk}H^k$ or $H^i=\displaystyle{\frac{1}{2}}\epsilon^{ijk}H_{jk}$. Contraction of both sides of (\ref{eq:B18}) with $\epsilon_{ijk}$, and using 
\begin{equation}
\frac{\delta\tilde{\Gamma}}{\delta F_{ij}}=\frac{1}{2}\epsilon^{ijk}
\frac{\delta\tilde{\Gamma}}{\delta B^k}\;,\label{eq:B20}
\end{equation}
results in
\begin{equation}
H_k=B_k+T\frac{\delta\tilde{\Gamma}}{\delta B^k}\;.\label{eq:B21}
\end{equation}
The thermodynamic potential $\Omega$ is given by
\begin{equation}
\tilde{\Gamma}=\beta\Omega\;.\label{eq:B22}
\end{equation}
If we define the thermodynamic potential density $\omega$ by 
\begin{equation}
\Omega=\int_{\Sigma}d\sigma_x\,\omega\;,\label{eq:B23}
\end{equation}
then we have
\begin{equation}
H_i=B_i-M_i\;,\label{eq:B24}
\end{equation}
where 
\begin{equation}
M_i=-\frac{\partial\omega}{\partial B^i}\label{eq:B25}
\end{equation}
defines the components of the magnetization.


\begin{thebibliography}{99}

\bibitem{bose24}
S.~N.~Bose, Z.~Phys. {\bf 26}, 178 (1924).

\bibitem{einsteinpreus24}
A.~Einstein, S.~B.~Preus~Akad.~Wiss., {\bf 22}, 261 (1924).

\bibitem{landaulifshitz69}
L.~D.~Landau and E.~M.~Lifshitz, {\em Statistical Physics\/},
(Pergammon, London, 1969).

\bibitem{may64}
R.~M.~May, Phys.~Rev.~A {\bf 135}, 1515 (1964).

\bibitem{kapusta81}
J.~I.~Kapusta, Phys.~Rev.~D {\bf 24}, 426 (1981).

\bibitem{haberweldon81}
H.~E.~Haber and H.~A.~Weldon, Phys.~Rev.~Lett. {\bf 46}, 1497 (1981).

\bibitem{haberweldon82a}
H.~E.~Haber and H.~A.~Weldon, Phys.~Rev.~D {\bf 25}, 502 (1982).

\bibitem{huang93}
K.~Huang, {\em Bose-Einstein condensation and superfluidity\/},
\newblock Technical Report MIT CTP 2218.

\bibitem{bensonbernsteindodelson91}
K.~M.~Benson, J.~Bernstein, and S.~Dodelson, Phys.~Rev.~D {\bf 44}, 2480 (1991).

\bibitem{bernsteindodelson91}
J.~Bernstein and S.~Dodelson, Phys.~Rev.~Lett. {\bf 66}, 683 (1991).

\bibitem{kirstentoms95}
K.~Kirsten and D.~J.~Toms, Phys.~Rev.~D {\bf 51}, 6886 (1995).

\bibitem{toms92}
D.~J.~Toms, Phys.~Rev.~Lett. {\bf 69}, 1152 (1992).

\bibitem{toms93}
D.~J.~Toms, Phys.~Rev.~D {\bf 47}, 2483 (1993).

\bibitem{alta}
M.~B.~Al'taie, J.~Phys.~A {\bf 11}, 1603 (1978).

\bibitem{singh}
S.~Singh and R.~K.~Pathria, J.~Phys.~A {\bf 17} (1984) 2983.

\bibitem{park}
L.~Parker and Y.~Zhang, Phys.~Rev.~D {\bf 44}, 2421 (1991).

\bibitem{cog}
G.~Cognola and L.~Vanzo, Phys.~Rev.~D {\bf 47}, 4575 (1993).

\bibitem{dowken}
J.~S.~Dowker and G.~Kennedy, J.~Phys.~A {\bf 11} 895 (1978).\\
J.~S.~Dowker and J.~P.~Schofield, Nucl.~Phys.~B {\bf 327}, 267 (1989).

\bibitem{bytvan}
A.~A.~Bytsenko, L.~Vanzo and S.~Zerbini, Phys.~Lett.~B {\bf 291}, 26 (1992).

\bibitem{kkthesis}
K.~Kirsten, Class.~Quantum Grav.~{\bf 8}, 2239 (1991).\\
K.~Kirsten, J.~Phys.~A {\bf 24}, 3281 (1991).

\bibitem{schafroth51}
M.~R.~Schafroth, Helv.~Phys.~Acta {\bf 24}, 645 (1951).

\bibitem{schafroth55}
M.~R.~Schafroth, Phys. Rev. {\bf 100}, 463 (1955).

\bibitem{may59}
R.~M.~May, Phys.~Rev. {\bf 115}, 254 (1959).

\bibitem{may65}
R.~M.~May, J. Math. Phys. {\bf 6}, 1462 (1965).

\bibitem{toms95}
D.~J.~Toms, Phys.~Rev.~D {\bf 51}, 1886 (1995).

\bibitem{toms95a}
D.~J.~Toms, Phys.~Lett.~B {\bf 343}, 259 (1995).

\bibitem{daicicfrankelkowalenko94}
J.~Daicic, N.~E.~Frankel, and V.~Kowalenko, Phys. Rev. Lett. {\bf 71}, 1779 (1993).

\bibitem{daicicfrankelgailiskowalenko94}
J.~Daicic, N.~E.~Frankel, R.~M.~Gallis, and V.~Kowalenko, Phys. Rep. {\bf 237}, 63 (1994).

\bibitem{elmforsperssonskagerstam93}
P.~Elmfors, D.~Persson, and B.~S.~Skagerstam, Phys.~Rev.~Lett. {\bf 71}, 480 (1993).

\bibitem{elmforsliljenbergperssonskagerstam95}
P.~Elmfors, P.~Liljenberg, D.~Persson, and B.~S.~Skagerstam, Phys.~Lett.~B {\bf 348}, 462 (1995).

\bibitem{daicic}
J.~Daicic and N.E.~Frankel, Phys.~Rev.~D {\bf 53}, 5745 (1996).

\bibitem{rojas}
H.~P.~Rojas, Phys.~Lett.~B {\bf 379}, 148 (1996).

\bibitem{physlett}
K.~Kirsten and D.~J.~Toms, Phys. Lett. B {\bf 368}, 119 (1996).

\bibitem{KirstenTomsPRA}
K.~Kirsten and D.~J.~Toms, Phys. Rev. A {\bf 54}, 4188 (1996).

\bibitem{bernard74}
C.~W.~Bernard, Phys.~Rev.~D {\bf 9}, 3312 (1974).

\bibitem{kapusta89}
J.~I.~Kapusta, {\em Finite Temperature Field Theory\/} (Cambridge University Press, Cambridge, 1989).

\bibitem{landsmanvanweert87}
N.~P.~Landsman and Ch.~G.~van~Weert, Phys.~Rep. {\bf 145}, 141 (1987).

\bibitem{toms94}
D.~J.~Toms, Phys.~Rev.~D {\bf 50}, 6457 (1994).

\bibitem{dewitt65}
B.~S.~DeWitt, {\em Dynamical Theory of Groups and Fields\/} (Gordon and Breach, New York, 1965).

\bibitem{hawking77}
S.~W.~Hawking, Commun.~Math.~Phys. {\bf 55}, 133 (1977).

\bibitem{critchleydowker76}
R.~Critchley and J.~S.~Dowker, Phys.~Rev.~D {\bf 13}, 3224 (1976).

\bibitem{TomsPRB}
D.~J.~Toms, Phys. Rev. B {\bf 50}, 3120 (1994).

\bibitem{Ziff}
R.~M.~Ziff, G.~E.~Uhlenbeck, and M.~Kac, Phys. Rep. {\bf 32C}, 169 (1977).

\bibitem{smithtoms}
J.~D.~Smith and D.~J.~Toms, Phys.~Rev.~D {\bf 53}, 5771 (1996).

\bibitem{gradshteynryzhik65}
I.~S.~Gradshteyn and I.~M.~Ryzhik, {\em Tables of Integrals, Series and Products\/} (Academic Press, New York, 1965).

\bibitem{KKDJTinprep}
K.~Kirsten and D.~J.~Toms, in preparation.

\bibitem{Hu}
D. J. O'Connor, B. L. Hu and T. C. Shen, Phys. Lett. B {\bf 130}, 31 
(1983);
B. L. Hu and D. J. O'Connor, Phys. Rev. Lett. {\bf 56}, 1613 (1986); 
ibid., Phys. Rev. D {\bf 36}, 1701 (1987).

\bibitem{Jackson}
J.~D.~Jackson, {\em Classical Electrodynamics\/} (Wiley, New York, 1962).

\bibitem{Gug}
E.~A.~Guggenheim, {\em Thermodynamics\/} (North-Holland, Amsterdam, 1959).

\bibitem{Schweber}
S.~S.~Schweber, {\em An Introduction to Relativistic Quantum Field Theory\/} (Row, Peterson and Company, Evanston, 1961).

\end{thebibliography}
\end{document}